\documentstyle[amsmath,amsfonts]{article}
\newcommand{\X}{\textrm{{\bf X}}}
\newcommand{\E}{\textrm{{\bf E}}}
\newcommand{\F}{\textrm{{\bf F}}}
\newcommand{\HH}{\textrm{{\bf H}}}

\newcommand{\cv}{{\cal V}}
\newcommand{\ch}{{\cal H}}
\newcommand{\cf}{{\cal F}}

\newcounter{sec}
\newtheorem{teo}[sec]{Theorem}
\newtheorem{lema}[sec]{Lemma}
\newtheorem{defi}[sec]{Definition}
\newtheorem{prop}[sec]{Proposition}

\newcommand{\al}{{\alpha}}
\newcommand{\la}{{\lambda}}

\begin{document}
%%%%%%%%%%%%%%%%%%%%%%%%%%%%%%%%%%%%%%%%%%%%%%%%%%%%%%%%%%%%%%%%5

\begin{titlepage}
\begin{center}
\renewcommand{\thefootnote}{\fnsymbol{footnote}}
{%\twelve
Centre de Physique Th\'eorique\footnote{Unit\'e Propre de
Recherche 7061 }, CNRS Luminy, Case 907}

{%\twelve
F-13288 Marseille -- Cedex 9}

\vspace{3 cm}

{\huge {Singular and non-singular eigenvectors\\ for the Gaudin model}}

\vspace{2 cm}

\setcounter{footnote}{0}
\renewcommand{\thefootnote}{\arabic{footnote}}

{\bf Daniela GARAJEU\footnote{\it garajeu@cpt.univ-mrs.fr}
and Annam\'aria KISS\footnote{\it kiss@cpt.univ-mrs.fr}}

\vspace{3 cm}

{\bf Abstract}\\

\vspace{0.5 cm}
\end{center}

We present a method to construct a basis of singular and non-singular common 
eigenvectors for Gaudin Hamiltonians
in a tensor product module of the Lie algebra $SL(2)$. 
The subset of singular vectors is completely described by analogy with covariant 
differential operators. The relation between singular eigenvectors and the Bethe Ansatz
is discussed. In each weight subspace the set of singular eigenvectors is completed to a basis, by a family of non-singular eigenvectors. 
We discuss also the generalization of this method to the case of an arbitrary Lie algebra.

\vfill

\noindent PACS: 03.65.-w, 02.20.Sv, 05.30.-d
\bigskip

\bigskip

\noindent {\it J. Math. Phys.} 42(8) 3497, 2001

\noindent CPT-2001/P.4148

\bigskip

\noindent anonymous ftp : ftp.cpt.univ-mrs.fr

\noindent web : www.cpt.univ-mrs.fr

\renewcommand{\thefootnote}{\fnsymbol{footnote}}

\end{titlepage}

\setcounter{footnote}{0}
\renewcommand{\thefootnote}{\arabic{footnote}}

%%%%%%%%%%%%%%%%%%%%%%%%%%%%%%%%%%%%%%%%%%%%%%%%%%%%%%%%%%%%%%%%

%%%%%%%%%%%%%%%%%%%%%%%%%%%%%%%%%%%%%%%%%%%%%%%%%%%%

\section{Introduction}

The Gaudin model \cite{Gaudinart} is an example of an integrable system in statistical quantum mechanics, 
associated to $N$ spin particles with magnetic interaction, based on  $N$ independent 
and commuting Hamiltonians. As originally formulated, this model is  
related to the Lie algebra $SL(2)$.
In this simplest case, 
the problem of the diagonalization of the Hamiltonians was partially solved. 
Using a constructive method, known as Bethe Ansatz, Gaudin \cite{Gaudin}
constructed a family of common eigenvectors for all Hamiltonians.
 
More generally, integrable systems can be associated to any semi-simple complex Lie algebra. 
For such systems, generalizations of the set of Gaudin Hamiltonians have been
constructed.\cite{FFR} 
The methods  proposed for diagonalization of these Hamiltonians revealed 
remarkable connections between integrable models and two 
dimensional conformal field theories. 
In this approach \cite{FFR}, 
the structure of the Bethe vectors for the Gaudin model is related to the 
representation theory of affine Lie algebras. The diagonalization 
of Gaudin Hamiltonians is based on a concept of invariant functionals  
(correlation functions) on tensor products of representations  of an affine Lie algebra 
at the critical level.

Another approach for the problem of diagonalization,\cite{BabujianFlume,Babujian} 
is related to the
connection between eigenvectors of the Gaudin Hamiltonians  and 
the  solutions \cite{VarchenkoSchechtman} of the Knizhnik-Zamolodchikov 
equations.\cite{KZ} In this approach, the common eigenvectors are 
constructed inductively and each eigenvector leads to an 
integral solution of the KZ equations.
Conversely, it was proved \cite{RV} that in the quasi-classical limit,  
the first term of the asymptotic solutions 
of KZ equation leads to a common eigenvector of the Hamiltonians. In this case the Bethe 
equations for the Gaudin model appear as conditions of critical points.

In section 2 we review some aspects concerning the Gaudin model, 
originally formulated for the Lie algebra $SL(2)$. 
We recall the expression of the $N$ Hamiltonians, introduced by Gaudin.  
The space $\Omega$ of physical states is defined as a tensor product of $N$ finite-dimensional 
highest weight representations of the Lie algebra $SL(2)$. 
It is decomposed as a direct sum of weight subspaces, {\it i.e.}, 
$\Omega =\oplus {\cal V}_m$. We explain how the Bethe Ansatz works to construct  
common eigenvectors for Gaudin Hamiltonians in each 
eigenspace ${\cal V}_m$. However, this method cannot give
all the common eigenvectors of Gaudin Hamiltonians. Therefore,
supplementary common eigenvectors have to be determined.

In section 3 we  give a general, recursive method to construct
a  basis of common eigenvectors in each invariant subspace ${\cal V}_m$.
Knowing a basis of ${\cal V}_{m-1}$, we construct a family of linear 
independent common eigenvectors in ${\cal V}_m$, which are nonsingular.
This family is completed to a basis of ${\cal V}_m$, by a basis of the 
subspace of singular vectors of ${\cal V}_m$.
Using an analog of the Gordan operator,\cite{Go} we give a basis of  Sing ${\cal V}_m$.
Finally, we show that the Bethe Ansatz gives a family of singular common eigenvectors 
for Gaudin Hamiltonians, but in some particular cases, this family could not be complete.

Section 4 is devoted to a discussion of the generalized model associated to an arbitrary 
simple Lie algebra.
We recall the generalization of the Gaudin Hamiltonians  
and explain how the Bethe Ansatz was generalized to 
construct common eigenvectors for these Hamiltonians. We show that 
Bethe equations, which appear as conditions that Bethe type vectors be 
common eigenvectors, are also conditions of singularity for them.
This constructive method does not ensure the completeness of the 
system of eigenvectors. 

%%%%%%%%%%%%%%%%%%%%%%%%%%%%%%%%%%%%%%%%%%%%%%%%%%%%

\section{Gaudin model and the $SL(2)$ Lie algebra}

The Gaudin spin model is related to the Lie algebra $SL(2)$, with generators
$E$, $F$ and $H$ satisfying the commutation relations:
\[
\left[E,F \right]=H;\qquad \left[H,E \right]=2E;\qquad \left[H,F \right]=-2F.
\]
For this Lie algebra we consider $N$ finite-dimensional highest weight modules:
$V_{\la_1}$,..., $V_{\la_N}$ with highest weights $\la_1$,..., $\la_N$
and highest weight vectors $v_{\la_1}$,...,  $v_{\la_N}$.
The tensor product of these $N$ modules constitutes the space of
physical states for a system of $N$ spin particles:
\[
\Omega= V_{\la_1}\otimes ...\otimes V_{\la_N}.
\]
The Lie algebra action on each vector $v=v_1\otimes ...\otimes
v_N$ of this tensor module is defined as
\begin{equation}
\X\, v\ =\ \sum_{i=1}^N X^{(i)} v,
\qquad \forall\ X\in SL(2),
\label{X}
\end{equation}
where $X^{(i)}$ denotes the operator on $\Omega$ which acts as $X$ on the
$i^{\hbox{\rm th}}$ module and as the identity operator on all other factors:
\begin{equation}
X^{(i)}\ v_1\otimes ...\otimes v_N=v_1\otimes ...\otimes Xv_i\otimes ...\otimes
v_N,
\qquad\forall\ v_1\otimes ...\otimes v_N\in\Omega\label{Ai}.
\end{equation}

For such a system of $N$ spin particles, Gaudin  proposed a set of $N$
Hamiltonians defined on $\Omega$, depending on $N$ distinct, complex
parameters $z_1,...,z_N$ :
\begin{equation}
{\cal H}_i(z_1,...,z_N)=\sum_{j=1,j\not= i}^N \frac{1}{z_i-z_j}
\left[\frac{1}{2} H^{(i)} H^{(j)}+E^{(i)} F^{(j)}+F^{(i)} E^{(j)}\right],\quad\forall\
i=1,...,N.
\label{Hi}
\end{equation}
All these operators commute:
\[
\left[{\cal H}_i,{\cal H}_j  \right] =0,\qquad \forall\ i,j=1,...,N,
\]
but they are not independent, because
$\sum_{i=1}^N {\cal H}_i=0.$
It can easily be verified that among the $N$ Gaudin Hamiltonians
there are exactly $N-1$ which are independent.
To integrate a system of $N$ spin particles, with $N$ degrees of freedom, the family
of commuting operators is completed by the Cartan generator
(also called the total spin operator), of which action on the tensor module is
\[
\HH\, v=\sum_{i=1}^N  H^{(i)} v,
\quad\forall\ v\in\Omega,
\]
and which commutes with all Gaudin Hamiltonians.

For this family of $N$ independent Hamiltonians which commute there is a
complete system of common eigenvectors in $\Omega$.
Our goal is to construct a basis in $\Omega$, of common eigenvectors of 
Gaudin Hamiltonians.

%----------------------------------------------------------------
\subsection{The structure of the space $\Omega$}
%----------------------------------------------------------------

In order to analyze the structure of $\Omega$ it is useful to recall some elements of 
the theory of highest weight
representations of the Lie algebra $SL(2)$. Such a representation
is completely determined by a highest weight vector $v_{\la}$, on which the
action of the algebra is given by
\[
H v_{\la}=\la v_{\la},\qquad E v_{\la}=0,
\]
and the representation space is generated by vectors:
\[
\{v_n=F^n\ v_{\la}\}_{n\in {\mathbb{N}}}\,.
\]
The action of the Lie algebra on these vectors is
\begin{eqnarray}
H v_{n}&=&\left(\la -2n\right) v_{n}\label{grad},\\
E v_{n}&=&n\left( \la -n+1\right) v_{n-1}\label{EFk},\\
F v_{n}&=&v_{n+1}\label{FFk}.
\end{eqnarray}

If the highest weight $\la$ is not a positive integer, the representation
is infinite dimensional and irreducible. If $\la\in {\mathbb{N}}$, then in
$\{v_n=F^n\ v_{\la}\}_{n\in {\mathbb{N}}}\ $ there is an invariant subspace, generated by
$\{v_{\la+1},\ v_{\la+2},...\}$ and the quotient representation is 
irreducible and finite dimensional, of dimension $\la +1$,
generated by vectors $\{v_n=F^n\ v_{\la}\}_{n=0,...,\la}$. 
We denote $V_{\la}$ this quotient representation, for which 
we have $F^n\ v_{\la}=0$ for all $n\geq \la+1$.

We will consider the space $\Omega$ as a tensor product of finite-dimensional
representations $V_{\la}$, which is completely determined by the vector
$v_0=v_{\la_1}\otimes ...\otimes v_{\la_N}$, called vacuum vector and is
generated by vectors  
%$\{v_{n_1...n_N}=
$\ \{F^{n_1}v_{\la_1}\otimes F^{n_2}v_{\la_2}...\otimes F^{n_N}v_{\la_N}
\}_{n_i=0,...,\la_i;\ i=1,...,N}$, with $n_i$
operators $F$ applied on the i$^{\mbox{\rm th}}$ component. Such a vector can be 
written as
a product of $m=n_1+n_2+...+n_N$ operators of type (\ref{Ai}), denoted
%\footnote{Such a simplified notation is useful for models with very 
%large number $N$ of particles ($N>>m$).}
\[
v_m^{(k_1,...,k_m)}=F^{(k_1)} F^{(k_2)}...F^{(k_m)} v_0,
\]
with $1\leq k_1\leq ...\leq k_m\leq N$.
Note that for finite-dimensional representations, $m$ can vary between 0 and a maximal value
$m_{\mbox{\tiny\rm max}}=\sum_{i=1}^N \la_i$.

Note also the particular action of the Hamiltonians on the vacuum vector:
\begin{equation}
\HH\, v_0=\left( \sum_{k=1}^N \la_k\right) v_0,\qquad
{\cal H}_i\, v_0=\left( \frac{1}{2}\sum_{j=1,j\not=i}^N
\frac{\la_i \la_j}{z_i -z_j}\right) v_0\label{ecvacuum}.
\end{equation}

The Cartan operator $\HH$ has a privileged place in the family of commuting
Hamiltonians. According to (\ref{grad}), it gives a gradation of the
representation spaces $V_{\la_i}$, which induces a gradation of the tensor
product module, on $\HH$-invariant subspaces:
\[
\Omega =\bigoplus\limits_{m=0}^{m_{\mbox{\tiny\rm max}}} {\cal V}_m,
\]
where ${\cal V}_m$ is a weight subspace, of weight $\sum_{i=1}^N \la_i -2m$,
also called space of spin deviation $m$. It is 
generated by $\HH$-eigenvectors:
\begin{equation}
\{v_m^{(k_1,...,k_m)}=F^{(k_1)}...F^{(k_m)}v_0\}_{1\leq k_1\leq...\leq k_m\leq N},
\label{bazaVm}
\end{equation}
which we call states of spin deviation $m$:
\[
\HH\ F^{(k_1)} F^{(k_2)}...F^{(k_m)} v_0=\left( \sum_{i=1}^N \la_i -2m \right)
F^{(k_1)} F^{(k_2)}...F^{(k_m)} v_0.
\]

Note that for spin deviations 
$m\leq\textrm{min}\left\{\la_1,...,\la_N\right\}$, a 
subspace ${\cal V}_m$ has the dimension
\begin{equation}
\mbox{\rm dim}{\cal V}_m=\sum_{1\leq k_1\leq ...\leq k_m\leq N}1=
\sum_{1\leq j_1< ...< j_m \leq N+m-1}1=C_{N+m-1}^m.
\label{dimVm}
\end{equation}

From the explicit form of the Gaudin Hamiltonians (\ref{Hi}) and 
from the action (\ref{grad})-(\ref{FFk}) of the $SL(2)$ Lie algebra 
on the spaces $V_{\la_i}$, 
it follows that the weight subspaces ${\cal V}_m$ are invariant under the 
action of any ${\cal H}_i$. Therefore, in each subspace ${\cal V}_m$,
we can construct a basis of common eigenvectors of ${\cal H}_i$.

%================================================================

\subsection{The construction of common eigenvectors by Bethe Ansatz}

Bethe Ansatz is a method to construct a family of common eigenvectors for 
Gaudin Hamiltonians in each invariant subspace ${\cal V}_m$, 
but this family is not a basis of ${\cal V}_m$.

Since  ${\cal V}_m$ is generated by (\ref{bazaVm}), any common eigenvector in 
${\cal V}_m$ has the form 
\[
\psi_m=\sum_{k_1=1}^{N}...\sum_{k_m=1}^{N} c_{k_1...k_m} F^{(k_1)}...F^{(k_m)} v_0.
\]
The central idea of the Bethe method is to consider the 
coefficients $c_{k_1...k_m}$ as rational complex functions:
\[
c_{k_1...k_m}=\frac{1}{w_1-z_{k_1}}...\frac{1}{w_m-z_{k_m}},
\]
depending on some unknown, distinct, complex parameters: $w_1,...,w_m$. 
We call such a vector a Bethe vector. Hereafter
we shortly present this method.

%-----------------------------

\subsubsection{Bethe vectors of spin deviation $m=1$}

The eigenspace ${\cal V}_1$ is generated by vectors
$\left\{F^{(k)}  v_0 \right\}_{k=1,...,N}$.
A Bethe vector of spin deviation $m=1$ is defined as an
expansion on this basis, with rational coefficients
depending on one complex parameter $w$:
\[
\psi_1(w)\ =\ \cf(w) v_0,
\]
where we denoted by $\cf(w)$ the operator  on $\Omega$:
\begin{equation}
{\cal F}(w)\ =\ \sum_{k=1}^{N}\frac{1}{w-z_k}F^{(k)}.
\label{calF}
\end{equation}
% Similarly, denoting: 
%\begin{equation}
%h(w)\ =\ \sum_{k=1}^{N}\frac{1}{w-z_k}H^{(k)}
%\label{h(w)}
%\end{equation}
Straightforward calculations give the commutator
\begin{equation}
\left[{\cal H}_i, {\cal F}(w)\right]\ =\ \cf(w)\frac{H^{(i)}}{(w-z_i)}-
\frac{F^{(i)}}{(w-z_i)}\sum_{k=1}^{N}\frac{H^{(k)}}{w-z_k}.
\label{comuthf}
\end{equation}
Applying this operator on $v_0$ and using (\ref{ecvacuum}) we obtain 
%\begin{equation}
%\left[{\cal H}_i, {\cal F}(w)\right]v_0=\frac{\la_i}{w-z_i}{\cal F}(w)v_0-
%\left( \sum_{k=1}^N \frac{\la_k}{w-z_k}\right) \frac{F^{(i)}}{w-z_i} v_0.
%\label{comut1}
%\end{equation}
%Then 
the action of a Gaudin Hamiltonian ${\cal H}_i$ 
on $\psi_1(w)$:
\[
{\cal H}_i\,\psi_1(w)=\left( \frac{1}{2}\sum_{j=1;j\not=i}^N\frac{\la_i\la_j}{z_i -z_j}+
\frac{\la_i}{w -z_i}\right) \psi_1(w) -
\left( \sum_{k=1}^N\frac{\la_k}{w -z_k}\right)\frac{F^{(i)}}{w-z_i}v_0,
\]
and we have the following lemma.
\begin{lema}\label{lema1}
Given $N$ distinct complex numbers $\left\{z_i \right\}_{i=1,...,N}$ and fixed, positive, 
integer highest weights $\left\{\la_i \right\}_{i=1,...,N}$,
the Bethe vector  $\psi_1(w)$ of spin deviation $m=1$ is an eigenvector for all Gaudin
Hamiltonians:
\[
{\cal H}_i\,\psi_1(w)=s_i^1\,\psi_1(w),\qquad\forall\ i=1,...,N,
\]
if the complex parameter $w$ satisfies the condition
\begin{equation}
\sum_{k=1}^N\frac{\la_k}{w -z_k}=0,
\label{bethe1}
\end{equation}
called Bethe equation associated to ${\cal V}_1$.
The eigenvalue $s_i^1$ of ${\cal H}_i$, depends on the solution
$w$ of this equation:
\[
s_i^1(w)=\frac{1}{2}\sum_{j=1;j\not=i}^N\frac{\la_i\la_j}{z_i -z_j}+\frac{\la_i}{w -z_i}.
\]
\end{lema}

\subsubsection{Bethe vectors of spin deviation $m=2$}

The eigenspace ${\cal V}_2$ is generated by vectors
$\left\{F^{(k_1)} F^{(k_2)} v_0 \right\}_{1\leq k_1\leq k_2\leq N}$.
A Bethe vector of spin deviation $m=2$ is defined
as an expansion on this basis, with rational coefficients
depending on two complex parameters $w_1,\ w_2$ :
\[
\psi_2(w_1,w_2)\ =\ {\cal F}(w_1){\cal F}(w_2)v_0.
\]
Remark that the order of the two operators $\cf$ is not significant because they commute.

In order to compute the action of a Hamiltonian on this state, we have to commute ${\cal H}_i$
successively with the two operators  ${\cal F}$.
Using (\ref{comuthf}) we obtain the commutator formula:
\begin{equation}
\left[\left[{\cal H}_i, {\cal F}(w_1)\right], {\cal F}(w_2)\right]
=\frac{2}{w_1-w_2}\left(
\frac{F^{(i)}}{w_1-z_i}{\cal F}(w_2)-
\frac{F^{(i)}}{w_2-z_i}{\cal F}(w_1)\right).
\label{comut2}
\end{equation}
From (\ref{comuthf}) and (\ref{comut2}) the action of a Gaudin Hamiltonian
${\cal H}_i$ on the state $\psi_2$ is
\begin{eqnarray}
{\cal H}_i\,\psi_2(w_1,w_2)&=&
\left( \frac{1}{2}\sum_{j=1;j\not=i}^N\frac{\la_i\la_j}{z_i -z_j}+
\frac{\la_i}{w_1 -z_i} + \frac{\la_i}{w_2 -z_i}\right) \psi_2(w_1,w_2) \nonumber\\
& &-\left(\sum_{k=1}^N\frac{\la_k}{w_1 -z_k}+
\frac{2}{w_2-w_1}\right)\frac{F^{(i)}}{w_1-z_i}\,\psi_1(w_2)\nonumber\\
& &-\left(\sum_{k=1}^N\frac{\la_k}{w_2 -z_k}+\frac{2}{w_1-w_2}\right)
\frac{F^{(i)}}{w_2-z_i}\,\psi_1(w_1),\nonumber
\end{eqnarray}
and we have the following lemma.
\begin{lema}
The  Bethe vector $\psi_2(w_1,w_2)$ of spin deviation $m=2$ is an eigenvector for all Gaudin
Hamiltonians:
\[
{\cal H}_i\,\psi_2(w_1,w_2)=s^2_i\,\psi_2(w_1,w_2),\qquad\forall\ i=1,...,N,
\]
if the complex parameters $w_1, w_2$ satisfy the Bethe equations associated to ${\cal V}_2$:
\begin{equation}
\sum_{k=1}^N\frac{\la_k}{w_1 -z_k}+\frac{2}{w_2-w_1}=0,\qquad
\sum_{k=1}^N\frac{\la_k}{w_2 -z_k}+\frac{2}{w_1-w_2}=0.\label{bethe2}
\end{equation}
The eigenvalues $s^2_i$ depend on the solutions $w_1, w_2$
of these equations:
\[
s^2_i(w_1,w_2)=\frac{1}{2}\sum_{j=1;j\not=i}^N\frac{\la_i\la_j}{z_i -z_j}+
\frac{\la_i}{w_1 -z_i}+\frac{\la_i}{w_2 -z_i}.
\]
\end{lema}

\subsubsection{Bethe vectors of spin deviation $m$}

In the subspace ${\cal V}_m$, generated by
vectors 
$\left\{F^{(k_1)}... F^{(k_m)} v_0 \right\}_{1\leq k_1\leq ...\leq k_m\leq N}$,
a Bethe vector
of spin deviation $m$ is defined as an
expansion with coefficients
depending on $m$ complex parameters $w_1,..., w_m$:
\begin{equation}
\psi_m(w_1,...,w_m) v_0\ =\ {\cal F}(w_1)...{\cal F}(w_m)v_0.
\label{bethevectpsim}
\end{equation}

The action of  a Gaudin Hamiltonian ${\cal H}_i$ on this state is calculated
by induction on $m$:
\begin{eqnarray}
{\cal H}_i\,\psi_m(w_1,...,w_m)v_0&=&{\cal H}_i\,{\cal F}(w_m)\psi_{m-1}(w_1,...,w_{m-1})v_0\nonumber\\
& &\hspace* {-2cm}={\cal F}(w_m){\cal H}_i\psi_{m-1}(w_1,...,w_{m-1})v_0+
\psi_{m-1}(w_1,...,w_{m-1})\left[{\cal H}_i\,,\,{\cal F}(w_m)\right] v_0\nonumber\\
& &\hspace* {-2cm}+
\left[\left[{\cal H}_i\,,\,{\cal F}(w_m)\right],\psi_{m-1}(w_1,...,w_{m-1})\right] v_0.\nonumber
\end{eqnarray}
The first term is computed from the induction hypothesis, the second from (\ref{comuthf})
and the last term from relation (\ref{comut2}). Putting all together it follows that
\begin{eqnarray}
{\cal H}_i\,\psi_m(w_1,...,w_m)v_0&=&
\left( \frac{1}{2}\sum_{j=1;j\not=i}^N\frac{\la_i\la_j}{z_i -z_j}+
\sum_{k=1}^m\frac{\la_i}{w_k -z_i}\right) \psi_m(w_1,...,w_m)v_0\nonumber\\
& &\label{Hipepsim}\\
& &\hspace* {-2cm}-\sum_{k=1}^m\left( \sum_{j=1}^N\frac{\la_j}{w_k -z_j}+
\sum_{l=1;l\not=k}^m\frac{2}{w_l-w_k}\right)
\frac{F^{(i)}}{w_k-z_i}
\psi_{m-1}(...,\hat{w_k},...) v_0,\nonumber
\end{eqnarray}
where $\psi_{m-1}(...,\hat{w_k},...)$ denotes $\psi_{m-1}\left( w_1,...,w_{k-1}, w_{k+1},...,w_m\right)$.
Hence, we obtain the following theorem.
\begin{teo}
The Bethe vector $\psi_m(w_1,...,w_m)v_0$ of spin deviation $m$ is a common eigenvector for all
Gaudin Hamiltonians:
\[
{\cal H}_i\,\psi_m(w_1,...,w_m)v_0=s^m_i\,\psi_m(w_1,...,w_m)v_0,\qquad\forall\ i=1,...,N,
\]
if the complex parameters $w_1,...,w_m$ satisfy the Bethe equations  associated to ${\cal V}_m$:
\begin{equation}
\sum_{j=1}^N\frac{\la_j}{w_k -z_j}+
\sum_{l=1;l\not=k}^m\frac{2}{w_l-w_k}=0,\qquad\forall\,k=1,...,m.
\label{bethem}
\end{equation}
The eigenvalues $s^m_i$ depend on the solution
of these equations:
\[
s_i^m(w_1,...,w_m)=\frac{1}{2}\sum_{j=1;j\not=i}^N\frac{\la_i\la_j}{z_i -z_j}+
\sum_{k=1}^m\frac{\la_i}{w_k -z_i}.
\]
\end{teo}

As observed in Ref. \cite{Eu2}, the Bethe Ansatz does not give
all the common eigenvectors of Gaudin Hamiltonians. Therefore,
supplementary common eigenvectors have to be determined, which will be done in the
following section. 

%==============================================================

\section{A construction of a basis of common eigenvectors for Gaudin Hamiltonians}

In this section we give a general method to construct
a  basis of common eigenvectors in each invariant subspace ${\cal V}_m$.
This is a recursive  method, which will be applied for 
subspaces ${\cal V}_m$, with $m\leq $ min$\{\la_1,...,\la_N\}$.
Knowing a basis of ${\cal V}_{m-1}$, we construct a family of linear independent 
common eigenvectors in ${\cal V}_m$, which are nonsingular. This family is 
completed to a basis of ${\cal V}_m$ by a basis of the subspace of singular 
vectors of ${\cal V}_m$.

There are two important properties of Gaudin Hamiltonians which are useful
in this section, namely they commute with the operators 
$\E$ and $\F$, on the tensor product module:
\begin{eqnarray}
\left[{\cal H}_i,\E\right]\ =\ 0\,,\label{HE}\\[2mm]
\left[{\cal H}_i,\F\right]\ =\ 0\,,\label{HF}
\end{eqnarray}
where $\E$ and $\F$ are defined by (\ref{X}) as
\[
\E=\sum_{i=1}^N E^{(i)},\qquad \F= \sum_{i=1}^N F^{(i)}.
\]
\begin{defi}
We call $v^s\in \Omega$ a singular vector of $\Omega$ if the generator $\E$
acts trivially on $v^s$,
\[
\E\, v^s\ =\ 0\,.
\]
We denote $\mbox{\rm Sing}\, V$ the subspace of singular vectors in $V$.
\end{defi}

It was shown in Ref. \cite{Etingof} that the dimension of the subspace Sing ${\cal V}_m$ is
\begin{equation}
\mbox{\rm dim}\left( \mbox{\rm Sing}\,{\cal V}_m\right)=C^m_{m+N-2}.
\label{dimSingVm}
\end{equation}
Since according to (\ref{dimVm}), for $m\leq\textrm{min}\{\la_1,...,\la_N\}$ 
the dimension of the space ${\cal V}_m$
is $C_{m+N-1}^m$, it follows that a basis of Sing ${\cal V}_m$ can be completed to 
a basis of ${\cal V}_m$, by a family of $C^{m-1}_{m+N-2}$ nonsingular linear 
independent vectors. The space spanned by this family of nonsingular  vectors is 
denoted NonSing ${\cal V}_m$. Hence,
\[
{\cal V}_m=\mbox{\rm Sing}\,{\cal V}_m\oplus \mbox{\rm NonSing}\,{\cal V}_m.
\]

From the property (\ref{HE}) we obtain the following.
\begin{lema}\label{lema5}
$\mbox{\rm Sing}\ {\cal V}_m$ is a $C_{m+N-2}^m$-dimensional vector subspace of
${\cal V}_m$ and it is ${\cal H}_i$-invariant:
\[
{\cal H}_i\ \left( \mbox{\rm Sing}{}{\cal V}_m\right) 
\subseteq \mbox{\rm Sing}{}{\cal V}_m,\quad\forall\, i=1,...,N.
\]
\end{lema}
Then $\mbox{\rm Sing}\,{\cal V}_m$ admits a basis formed by  singular common
eigenvectors of Gaudin Hamiltonians ${\cal H}_i$. 
Denote ${\cal B}_m^s$ this basis.

From the property (\ref{HF}) of the Gaudin Hamiltonians we can construct 
recursively a basis of common eigenvectors in NonSing ${\cal V}_m$:
\begin{prop}
The basis of common eigenvectors in $\mbox{\rm NonSing}\,{\cal V}_m$ is obtained
by the application of the operator $\F$ on all common eigenvectors (singular
and nonsingular) which form the basis of the invariant subspace ${\cal V}_{m-1}$.
Denote ${\cal B}_m^{ns}$ this basis:
\[
{\cal B}_m^{ns}=\F\left( {\cal B}_{m-1}\right),
\]
where $ {\cal B}_{m-1}$ is a basis of common eigenvectors of ${\cal V}_{m-1}$.
\label{prop6}
\end{prop}

{\bf Proof:} (1) The vectors of ${\cal B}_m^{ns}$ are linear independent. 
Indeed, for $m\leq\textrm{min}\{\la_1,...,\la_N\}$ the operator 
$\F:{\cal V}_{m-1}\longrightarrow {\cal V}_m$ is injective, since
$\textrm{Ker}\,\F=\{0\}$. Moreover, the injectivity of $\F$ implies that the number 
of elements of the family 
${\cal B}_m^{ns}$ is equal to the dimension of  ${\cal V}_{m-1}$, namely 
$C^{m-1}_{m+N-2}$.

(2) The vectors of ${\cal B}_m^{ns}$ are common eigenvectors of ${\cal H}_i$. 
If $v_{m-1}\in {\cal B}_{m-1}$ is a common eigenvector:
\[
{\cal H}_{i}\,v_{m-1}\ =\ a_{m-1}^i v_{m-1},
\]
then, from the property (\ref{HF}) it follows that the vector 
$v_m=\F\, v_{m-1}\in {\cal B}_m^{ns}$ is also an eigenvector, with
the same eigenvalue $a_{m-1}^i$.

(3) The vectors of ${\cal B}_m^{ns}$ are nonsingular.\\
Note first that in ${\cal V}_{0}$ there is only one vector, the vacuum
vector $v_0$. Thus, there is one common eigenvector $v_1=\F\,
v_0$ in the family ${\cal B}_{1}^{ns}$, which is indeed
nonsingular:
\[
\E\, v_1\ =\ \E\,\F\, v_0\ =\ \HH\, v_0\ =\, \left( \sum_{i=1}^N\la_i\right)  v_0\not=0.
\]

Now suppose by induction that for all $k\leq m-1$, with 
$m\leq\textrm{min}\{\la_1,...,\la_N\}$, we have constructed
a basis of ${\cal V}_k$ of the form ${\cal B}_k^{s}\cup {\cal B}_k^{ns}$,
where ${\cal B}_k^{ns}$ is obtained by application of the operator
$\F$ on vectors of the basis ${\cal B}_{k-1}$ of ${\cal V}_{k-1}$. 
Then for a vector $v_{m}^{ns}$ of ${\cal B}_{m}^{ns}$
we have
\begin{equation}
\E\, v_{m}^{ns} = \E\,\F\, v_{m-1} = \HH\, v_{m-1}+\F\,\E\, v_{m-1}
= \left(\sum_{i=1}^N\la_i-2(m-1)\right)v_{m-1}+\F\,\E\, v_{m-1}.
\label{non-sing}
\end{equation}
If $v_{m-1}$ is singular, then the last term of (\ref{non-sing}) vanishes and
 $v_{m}^{ns}$ is nonsingular, since 
\[
\sum_{i=1}^N\la_i-2(m-1)\geq Nm-2(m-1)>0,\qquad\forall\, N\geq2.
\]
If $v_{m-1}$ is nonsingular, then by the induction hypothesis, there is a 
$k\in \{1,...,m-1\}$ such that 
$v_{m-1}=\F^k\, v_{m-1-k}^s$ with $v_{m-1-k}^s$ a common singular eigenvector
of ${\cal B}_{m-1-k}^s$. Then relation (\ref{non-sing}) becomes
\begin{eqnarray}
\E\, v_{m}^{ns}&=&(k+1)\left(\sum_{i=1}^N\la_i-2(m-1)+k\right)v_{m-1}\not=0,
\qquad\forall\, N\geq2.
\label{colin}
\end{eqnarray}

4) The vectors of ${\cal B}_m^{ns}$ are linear independent of 
the singular vectors of the basis ${\cal B}_m^s$.\\
Consider a null linear combination of the vectors $v_m^{ns}(i)\in{\cal B}_m^{ns}$
and  $v_m^{s}(j)\in{\cal B}_m^s$:
\[
\sum_i a(i)\, v_m^{ns}(i)+\sum_j b(j)\, v_m^{s}(j)=0.
\]
Applying the operator $\E$, the second sum (of singular vectors) vanishes. Hence,
\[
\sum_i a(i)\,\E\, v_m^{ns}(i)=0.
\]
It follows from (\ref{colin}) that each vector $\E\, v_m^{ns}(i)$ is colinear with the 
vector $v_{m-1}(i)$, which form the basis of ${\cal V}_{m-1}$. Then, the 
coefficients $a(i)$ are all zero. Since $v_m^{s}(j)$ form also a basis, 
in Sing ${\cal V}_m$,  the coefficients $b(j)$ vanish too.

Properties (1)-(4) from above show that ${\cal B}_m^{ns}=\F\left( {\cal B}_{m-1}\right)$ 
is a basis of NonSing ${\cal V}_m$ and then ${\cal B}_m^{ns}\cup {\cal B}_{m}^s$ 
is a basis of common eigenvectors in ${\cal V}_m$.
In this basis, the sub-family ${\cal B}_m^{ns}$ is completely determined by 
Proposition \ref{prop6}, whereas for the subset of singular vectors ${\cal B}_{m}^s$
we have only the existence Lemma \ref{lema5} and not the structure. 
In the next section we will characterize the singular vectors of a tensor product
module of $SL(2)$.

%========================================================

\subsection{Singular vectors of the $SL(2)$ tensor modules}

For a finite-dimensional highest weight module $V_\la$ of the Lie
algebra $SL(2)$ there is only one singular vector: the highest weight vector
$v_\la$. Nevertheless, for tensor product modules
$\displaystyle{\Omega=\overset{N}{\underset{i=1}{\otimes}}V_{\la_i}}$ the vacuum vector
$v_0=\overset{N}{\underset{i=1}{\otimes}}v_{\la_i}$ is not the only singular vector. There exist
singular vectors in every subspace ${\cal V}_m$.

\subsubsection{The case $N=2$}

Consider first $N=2$ and $\Omega=V_{\la_1}\otimes V_{\la_2}\ $ with the
decomposition on invariant subspaces 
$\displaystyle{\Omega=\overset{m_{max}}{\underset{m=0}{\oplus}}{\cal V}_m}$.
In this case an invariant subspace ${\cal{V}}_m$ is generated by vectors:
\[
\left\{ F^{k}v_{\la_1}\otimes F^{m-k}v_{\la_2}\right\}_{k=0,...,m},\, 
\]
and the singular vectors of such a subspace are characterized by the following
proposition.
\begin{prop}
Let ${\cal V}_m$ be an invariant subspace of $\ \Omega=V_{\la_1}\otimes V_{\la_2}$,
with $\ m\leq \textrm{min}\left\{\la_1,\la_2\right\}$. Then a vector
$v_m^s\in{\cal{V}}_m$,
\[
v_m^s\ =\ \sum_{k=0}^m c_k\ F^kv_{\la_1}\otimes F^{m-k}v_{\la_2},
\]
is singular if and only if the coefficients $c_k$ satisfy the conditions
\begin{equation}
c_{k+1}(k+1)(k-\la_1)+c_k\,(m-k)(m-k-1-\la_2) = 0,\qquad \forall\ k=0,...,m-1.
\label{cond_ck}
\end{equation}
\end{prop}
The proof is based on straightforward calculation using relation (\ref{EFk}):
\begin{eqnarray}
\E\, v_m^s&=&E^{(1)}\,v_m^s+E^{(2)}\,v_m^s\nonumber\\
%&=&\sum_{k=0}^m c_k k(\la_1-k+1)F^{k-1}v_{\la_1}\otimes F^{m-k}v_{\la_2}\nonumber\\
%&&+\sum_{k=0}^m c_k (m-k)(\la_2-m+k+1)F^kv_{\la_1}\otimes F^{m-k-1}v_{\la_2}\nonumber\\[2mm]
&=&\sum_{k=0}^{m-1}\left\{c_{k+1}(k+1)(\la_1-k)+c_k(m-k)(\la_2+k+1-m)\right\}
F^kv_{\la_1}\otimes F^{m-k-1}v_{\la_2}\nonumber.
\end{eqnarray}

Remark that the conditions (\ref{cond_ck}), satisfied by the coefficients $c_k$, 
coincide with the conditions determined in Ref.  \cite{Eu1}, 
satisfied by the coefficients of a bilinear
differential operator which is projective covariant. The system
(\ref{cond_ck}) was solved in this article and for $m\leq
\textrm{min}\left\{\la_1,\la_2\right\}$ it admits a unique solution (up to
a constant factor):
\[
c_k = (-1)^k\, C_m^k\,\frac{(m-k-\la_2)_k}{(-\la_1)_k},\qquad\forall\, k=0,...,m,
\]
with $C_m^k$ the binomial coefficient and $(x)_i$ the Pochhammer symbol:
\[
(x)_i=x(x+1)...(x+i-1),\ \forall\, i\in{\mathbb{N}}^* ,\qquad (x)_0=1.
\]
By analogy with covariant differential operators, 
we introduce the bilinear operator $P_m$ defined on the 
subspace ${\cal V}_m$ of $\Omega=V_{\la_1}\otimes V_{\la_2}$ by
\begin{equation}
P_m(v_1\otimes v_2)=\sum_{k=0}^m\, (-1)^k\, C_m^k\,
\frac{(m-k-\la_2)_k}{(-\la_1)_k}\,
F^k v_1\otimes F^{m-k} v_2,
\label{bil_op}
\end{equation}
which is analogous to the Gordan operator.

{\bf Conclusion.} In the case of $N=2$ in every invariant subspace 
${\cal V}_m$ with $m\leq
\textrm{min}\left\{\la_1,\la_2\right\}$ there
is a unique singular vector (up to a constant factor) which is
\[
v_m^s = P_m(v_{\la_1}\otimes v_{\la_2}) = \sum_{k=0}^m\, (-1)^k\,
C_m^k\,\frac{(m-k-\la_2)_k}{(-\la_1)_k}\, F^kv_{\la_1}\otimes F^{m-k}v_{\la_2}\,.
\]

\subsubsection{The case $N=3$}

Consider now $N=3$ and $\Omega=V_{\la_1}\otimes V_{\la_2}\otimes V_{\la_3}$ with
its decomposition on invariant subspaces. 
In this case, a subspace ${\cal V}_m$ is generated by vectors:
\[
\left\{ F^{k_1}v_{\la_1}\otimes F^{k_2}v_{\la_2}\otimes
F^{m-k_1-k_2}v_{\la_3}\right\}_{\mbox{\scriptsize{$\begin{array}{l}
k_1=0,...,m\\k_2=0,...,m-k_1\end{array}$}}},
\]
and the singular vectors of this subspace are characterized by the following
result.
\begin{prop}
Let ${\cal V}_m$ be an invariant subspace of $\ \Omega=V_{\la_1}\otimes
V_{\la_2}\otimes V_{\la_3}$, with $\ m\leq
\textrm{min}\left\{\la_1,\la_2,\la_3\right\}$. Then a vector
$v_m^s$ of ${\cal V}_m$,
\begin{equation}
v_m^s\ =\ \sum_{k_1=0}^m \sum_{k_2=0}^{m-k_1}\,c_{k_1k_2}\ 
F^{k_1}v_{\la_1}\otimes F^{k_2}v_{\la_2}\otimes F^{m-k_1-k_2}v_{\la_3},
\label{vect_N3}
\end{equation}
is singular if and only if the coefficients $c_{k_1k_2}$ satisfy the
conditions:
\begin{eqnarray}
& &c_{k_1+1,k_2}(k_1+1)(k_1-\la_1)
+c_{k_1,k_2+1}(k_2+1)(k_2-\la_2)\nonumber\\
& &+c_{k_1,k_2}(m-k_1-k_2)(m-k_1-k_2-1-\la_3)= 0,
\label{cond_ck12}\\
& &\forall\, k_1=0,...,m-1,\ \forall\, k_2=0,...,m-1-k_1\nonumber.
\end{eqnarray}
\end{prop}
The proof is analogous to that of $N=2$. As in the case $N=2$, we note that the
system (\ref{cond_ck12}), which has to be satisfied by the coefficients of the
development (\ref{vect_N3}) in order that $v_m^s$ be singular, coincides with
the conditions satisfied by the coefficients of a trilinear
differential operator which is projective covariant \cite{Eu1}. The system
(\ref{cond_ck12}) was also solved in Ref. \cite{Eu1} and for $m\leq
\textrm{min}\left\{\la_1,\la_2,\la_3\right\}$ it admits $m+1$ linear
independent solutions. In addition, it was shown that the space of covariant
trilinear operators is generated only by successive applications of covariant
bilinear operators.

%These results allow us to formulate the following conclusion concerning
%singular vectors in the case $N=3$:

{\bf Conclusion.} In the case of $N=3$ in every subspace ${\cal
V}_m$ with $m\leq
\textrm{min}\left\{\la_1,\la_2,\la_3\right\}$ there
are $m+1$ linear independent singular vectors:
\[
\textrm{dim}\,(\mbox{\rm Sing}\,{\cal V}_m)\ =\ m+1.
\]
Moreover,  using the operator on $\Omega$ introduced in (\ref{Pgras}) 
we can construct a basis of Sing ${\cal V}_m$, given by the vectors
\[
\left\{\,v_m^s\, =\, {\mathbf{P}}_{3,k}\,
\left({\mathbf{P}}_{2,m-k}\,(v_{\la_1}\otimes v_{\la_2})\otimes v_{\la_3}
\right)\,\right\}_{k=0,...,m}.
\]

\subsubsection{General case}

The results obtained for $N=2$ and $N=3$ can be generalized for an arbitrary
$N$. For each $N\geq 2$, consider the space 
$\Omega=\Omega^N=\Omega^{N-1}\otimes {\cal V}_{\la_N}$, where 
$\Omega^{N-1}=\overset{N-1}{\underset{i=1}{\otimes}}{\cal V}_{\la_i}$ and define on $\Omega^N$
the operator
\begin{equation}
{\mathbf{P}}_{N,m}(v_1\otimes v_2)=\sum_{k=0}^m\, (-1)^k\, C_m^k\,
\frac{(m-k-\la_2)_k}{(-\la_1)_k}\,
\F^k v_1\otimes F^{m-k} v_2,
\quad\forall\,v_1\in\Omega^{N-1},\,v_2\in{\cal V}_{\la_N},
\label{Pgras}
\end{equation}
where $\F=\sum_{i=1}^{N-1} F^{(i)}$ is defined by (\ref{X}) on $\Omega^{N-1}$.
Then we have the following.
\begin{prop}
Let ${\cal V}_m$ be an invariant subspace of 
$\ \Omega^N $ 
with $m\leq \textrm{min}\left\{\la_1,...,\la_N\right\}$. Then in ${\cal V}_m$ there is a
family of $C_{m+N-2}^m$ linear independent singular vectors, 
given by the formula
\begin{equation}
v_m^s\ =\ {\mathbf{P}}_{N,k}\left(v_{m-k}^s\otimes v_{\la_N}\right),\ 
k=0,...,m,
\label{vect_N}
\end{equation}
where $v_{m-k}^s$ is a singular vector  of
weight $\displaystyle{\sum_{i=1}^{N-1}\la_i-2(m-k)}$ in the basis of the subspace
$\mbox{\rm Sing}\,{\cal V}_{m-k}$ of $\ \Omega^{N-1}$.
\label{prop9}
\end{prop}

The singularity of these vectors arises from straightforward calculation 
analogous to the case $N=2$.

These vectors are linear independent because $v_{m-k}^s$ are the 
elements of the basis of the subspace
Sing ${\cal V}_{m-k}$ of $\Omega^{N-1}$ and 
for different $k$ the maximal number of operators $F$ applied on the last
component $v_{\la_N}$ is different.

The fact that by this construction we obtain exactly $C_{m+N-2}^m$ linear
independent singular vectors, can be demonstrated by induction with respect to
$N$: For $N=2$ and $N=3$ this number of singular vectors was already obtained.
Suppose now that for an arbitrary $N$ the number of linear independent
singular vectors $v_{m-k}^s$ in ${\cal V}_{m-k}$ is
$C_{m-k+N-2}^{m-k}$. Then for $N+1$, the number of linear independent singular
vectors obtained by the construction (\ref{vect_N}) is
\[
\sum_{k=0}^m C_{m-k+N-2}^{m-k}\ =\ \sum_{j=0}^m C_{j+N-2}^{j}
\ =\ C_{m+N-1}^m\,.
\]

Proposition \ref{prop9} allows to construct inductively a basis in Sing ${\cal V}_m$
which has the form
\begin{equation}
\left\{{\mathbf{P}}_{N,k_{N-1}}\left(...{\mathbf{P}}_{3,k_2}\left( 
{\mathbf{P}}_{2,k_1}(v_{\la_1}
\otimes v_{\la_2})
\otimes v_{\la_3}\right)...\otimes v_{\la_N}\right)\right\}_
{\mbox{\scriptsize{$
\begin{array}{l} k_i=0,...,m\\k_1+k_2+...+k_{N-1}=m\end{array}$}}}.
\label{bazasing}
\end{equation}

{\bf Remark.}  
We point out that in this section we have considered invariant 
subspaces ${\cal V}_m$ of $\Omega$ 
with spin deviation $m$ which does not exceed any of the $N$ weights 
$\la_i\in{\mathbb{N}}$:  
\[
m\leq \textrm{min}\left\{\la_1,...,\la_N\right\}.
\]
If $m$ is greater than at least one of the weights $\la_i$, then in the set 
\[
\left\{ F^{n_1}v_{\la_1}\otimes...\otimes  F^{n_N}v_{\la_N}\right\}_{n_1+...+n_N=m},
\]
there are elements for which the number of operators $F$ acting on $v_{\la_i}$ 
is greater then $\la_i$. These vectors are zero because for finite-dimensional irreducible 
representations of $SL(2)$ we have 
$\ F^nv_\la=0,\ \forall\, n\geq\la+1$. 
Therefore in this case the dimension of the space $\cv_m$ is less then $C^m_{m+N-1}$ 
and depends on the weights $\la$ which are less then $m$.

Moreover, if $m$ exceeds one or more of the weights $\la$, the dimension of the 
space Sing ${\cal V}_m$ is also less 
than $C^m_{m+N-2}$. For example, in the case $N=2$, it was shown in Ref. \cite{Eu1} that 
for $m\leq m_{max}=\la_1+\la_2$, but $m>\la_1$ and $m>\la_2$ there are  
two independent bilinear covariant differential operators, to which correspond singular 
vectors of the type
\[
P_{m-(1+\la_1)}\left( F^{1+\la_1}v_{\la_1}\otimes v_{\la_2}\right)
\textrm{ and }P_{m-(1+\la_2)}\left( v_{\la_1}\otimes F^{1+\la_2} v_{\la_2}\right),
\]
which are zero in the considered finite-dimensional irreducible representations of $SL(2)$. 
For an arbitrary $N$ we claim that relation (\ref{bazasing}) still gives the basis of 
Sing ${\cal V}_m$ but some of the elements of the form (\ref{bazasing}) vanish.
This fact implies the decreasing of the dimension of Sing ${\cal V}_m$.

%%%%%%%%%%%%%%%%%%%%%%%%%%%%%%%%%%%%%%%%%%%%%%%%%%%%%%%%%%%%%

\subsection{Singular common eigenvectors and Bethe Ansatz}

The basis (\ref{bazasing}) of Sing ${\cal V}_m$ constructed in the previous section 
is not a basis of common eigenvectors. In this section we show that 
the Bethe eigenvectors constructed in Sec. II using the Bethe Ansatz are singular, 
but their number could be less than the dimension of  Sing ${\cal V}_m$.
More precisely, the Bethe equations (\ref{bethem}), which
appear in the Bethe Ansatz as conditions that Bethe type vectors
(\ref{bethevectpsim}) be common eigenvectors of $\ch_i$ are 
also conditions that Bethe type vectors be singular.

\begin{prop}
If the distinct parameters $w_1,...,w_m$ are solutions of the Bethe system:
\begin{equation}
\sum_{j=1}^N\frac{\la_j}{w_k -z_j}+
\sum_{l=1;l\not=k}^m\frac{2}{w_l-w_k}=0,\qquad\forall\,k=1,...,m,
\label{eqprop10}
\end{equation}
then the corresponding Bethe type vector 
$\psi_m(w_1,...,w_m)\,v_0=
{\cal F}(w_1)...{\cal F}(w_m)v_0$ of ${\cal V}_m$ is a singular vector
of the tensor product module $\Omega$.
\end{prop}

{\bf Proof.} Using the commutators,
\[
\left[\E,\cf(w)\right]=h(w)\quad\textrm{ and }\quad
\left[h(w_1),\cf(w_2)\right]=
\frac{2}{w_1-w_2}\left(\cf(w_1)-\cf(w_2)\right),\nonumber
\]
with $h(w)=\sum_{k=1}^{N}\frac{1}{w-z_k}H^{(k)}$, we obtain
\[
\E\,\psi_m(w_1,...,w_m)\,v_0\,=\sum_{i=1}^m\left(\sum_{k=1}^N\frac{\la_k}{w_i-z_k}
+\sum_{\mbox{\scriptsize{$\begin{array}{l}k=1\\k\not=i\end{array}$}}}^m
\frac{2}{w_k-w_i}\right)\psi_{m-1}(w_1,...,w_{i-1},w_{i+1},...,w_m)\,v_0.
\]
Therefore if the parameters $w_1,...,w_m$ are solutions of the system
(\ref{eqprop10}), then $\psi_m(w_1,...,w_m)$ is a singular vector.

The number of common Bethe eigenvectors is equal to the 
number of distinct solutions of the Bethe system (\ref{eqprop10}).
If they are $C_{m+N-2}^m$
then they form a basis of Sing ${\cal V}_m$. 
For $m=1$ and $m=2$ we can easily show that the number of complex solutions of the 
Bethe system is $N-1$ and $N(N-1)/2$, respectively, but we cannot say how many of 
them are distinct. For example, in the case $m=1$, $N=3$, the Bethe equation is 
equivalent to a second order equation. For some particular values of the parameters 
$\la_i$ and $z_i$, $i=1,2,3$, this equation could have a double solution.

For arbitrary $m$ and $N$, the number of solutions of the Bethe system
seems to be $C_{m+N-2}^m$ but to our knowledge there is no estimate of 
the number of distinct solutions. In any case, the Bethe Ansatz gives a family 
of singular common eigenvectors which could be completed, if necessary, to 
a basis of Sing ${\cal V}_m$ with common eigenvectors constructed 
using the basis  (\ref{bazasing}).

%===========================================================================

\section{The Gaudin model for an arbitrary simple Lie algebra $G$}

Consider a simple Lie algebra $G$, of dimension $d$ and rank $r$.
Denote $\Delta$ the root system of $G$, $\Delta_+$ the system of positive roots
and  $\Delta_0$ the system of simple roots. The Cartan basis of $G$ is formed by
the Cartan generators $\left\{H_i \right\}_{i=1,...,r}$, the generators of positive
roots $\left\{E_{\alpha}\right\}_{\alpha\in\Delta_+}$ and the generators of negative
roots $\left\{F_{\alpha}=E_{-\alpha}\right\}_{\alpha\in\Delta_+}$.
The commutation relations are
\begin{eqnarray}
& &\left[E_{\al},F_{\al} \right]=\frac{2}{<\al,\al>}H_{\al};\qquad \left[H,E_{\al} \right]=\al(H) E_{\al};
\qquad \left[H,F_{\al} \right]=-\al(H) F_{\al};\nonumber\\
& &\left[E_{\al},E_{\beta} \right]=N_{\al,\beta}E_{\al +\beta},
\qquad\forall\ \al,\beta\in\Delta,\ \mbox{\rm such that}\ \al +\beta\in\Delta.\nonumber
\end{eqnarray}

The Killing form of $G$ defines a symmetrical, $G$-invariant,
bilinear form $<\,,\,>$ on $G$, which allows us to identify the
Cartan sub-algebra ${\cal K}$ of $G$ with its dual, by the
isomorphism $\alpha(H)=<H_{\alpha},H>$, for all $H\in
{\cal K}$. The scalar product on ${\cal K}$ induces a scalar product on  its
dual, which is nondegenerate,
\[
<\alpha,\beta>=<H_{\alpha},H_{\beta}>,
\]
normalized  such that
$<\alpha_l,\al_l>=2$ for a long root $\al_l$.

All generators of $G$ are orthogonal with respect to
the bilinear form $<\, ,\,>$ except
\begin{eqnarray}
<H_{i},H_{i}>&=&\frac{1}{2}\chi\ ,\ \forall\ i=1,...,r,\nonumber\\
<E_{\alpha},F_{\alpha}>&=&\frac{2}{<\al,\al>} ,\ \forall\ \al\in\Delta_+,\nonumber
\end{eqnarray}
where $\chi=\sum_{i=1}^r\al^2_{l}(H_i)$ is the square of the length of a long root.
Then, for the Cartan basis we can define the dual basis with respect to this
bilinear form:
\[
\tilde{H}_i=\frac{2}{\chi}H_i;\qquad
\tilde{E}_{\al}=\frac{<\al,\al>}{2}F_{\alpha};\qquad
\tilde{F}_{\al}=\frac{<\al,\al>}{2}E_{\alpha},
\]
such that $<H_i,\tilde{H}_j>=\delta_{ij}$ and
$<E_{\al},\tilde{E}_{\beta}>=<F_{\al},\tilde{F}_{\beta}>=\delta_{\al\beta}$.

For this algebra, consider $N$ finite-dimensional highest weight $G$-modules, 
$\ V_{\la_1}$,..., $V_{\la_N}$, with  dominant integral highest weights
$\la_1$,..., $\la_N$ and  highest weight vectors $v_{\la_1}$,...,  $v_{\la_N}$.
The tensor product module of these $G$-modules is the space of physical states:
\[
\Omega = V_{\la_1}\otimes ...\otimes V_{\la_N}.
\]
Recall that the Lie algebra action on this tensor module is defined by (\ref{X}).

 The fundamental building blocks of the generalized Gaudin Hamiltonians 
are the qua\-dra\-tic generalized Casimir operators,
%\cite{Kac83}, 
defined on $\Omega$ as
\begin{equation}
\omega^{(ij)}\ =\ \sum_{a=1}^d I_a^{(i)}\,\tilde{I}_a^{(j)},\qquad\forall\, i,j=1,...,N,
\label{casimir}
\end{equation}
where $\left\{I_a\right\}_{a=1,d}$ is a basis of $G$ and 
$\left\{\tilde{I}_a\right\}_{a=1,d}$ its dual with respect to the scalar product,  
$<I_a,\tilde{I}_b>=\delta_{ba}$.
We recall two fundamental properties of these operators, 
which will be useful in this section.

(1) $\omega^{(ij)}$ are independent of the choice of the basis in $G$. 
In particular,  in the Cartan basis they take the form
\[
\omega^{(ij)}\ =\ \frac{2}{\chi}\sum_{l=1}^r H_l^{(i)} H_l^{(j)}+
\sum_{\al\in\Delta_+}\!\frac{<\al,\al>}{2}\left( E_{\al}^{(i)} F_{\al}^{(j)}+ 
F_{\al}^{(i)} E_{\al}^{(j)}\right);
\]

(2) $\omega^{(ij)}$ commute with the action of the algebra on the tensor
module:
\[
\left[\omega^{(ij)},\X \right]=0,\ \forall X\,\in G.
\]

The operators $\omega^{(ij)}$ do not commute between themselves, but we can 
construct $N$ linear combinations which commute.
\begin{lema}
The operators $ \omega_i:\Omega\longrightarrow\Omega,\ i=1,...,N$,
\[
\omega_i\ =\ \sum_{j=1,j\not=i}^N c_{ij}\, \omega^{(ij)},
\]
commute: $\ [\omega_i,\omega_j]=0,\quad\forall i,j=1,...,N\ $ 
if and only if the coefficients $c_{ij}$ satisfy the equations
\begin{equation}
c_{ik}\,c_{jk}-c_{ik}\,c_{ji}-c_{jk}\,c_{ij}\ =\ 0,\quad\forall\, i\not=j\not=k\not=i,\ 
i,j,k=1,...,N.
\label{eqc}
\end{equation}
\end{lema}

In particular, if $z_1,...,z_N$ are distinct complex parameters, 
the coefficients $c_{ij}=\frac{1}{z_i-z_j}$ fulfill the conditions (\ref{eqc}).
With these coefficients we can construct the following $N$ operators on $\Omega$:
\begin{equation}
{\cal H}_i=\sum_{i=1,i\not=j}^N\frac{1}{z_i-z_j}\,\omega^{(ij)},
\label{Higen}
\end{equation}
which are the generalized Gaudin Hamiltonians.
They preserve the properties (1) and (2) of the operators $\omega^{(ij)}$. 
More precisely, if we consider the Cartan basis in $G$, we have
\begin{eqnarray}
\left[{\cal H}_i,\HH_{\al}\right]&=&0, \label{P1}\\
\left[{\cal H}_i,\E_{\al}\right]&=&0 \label{P2},
\qquad\forall\,i=1,...,N;\ \forall\,\al\in\Delta_+,\\
\left[{\cal H}_i,\F_{\al}\right]&=&0, \label{P3}
\end{eqnarray}
where $\ \HH_{\al}=\sum_{i=1}^N H_{\al}^{(i)}$, $\ \E_{\al}=\sum_{i=1}^N E_{\al}^{(i)}$, 
$\ \F_{\al}=\sum_{i=1}^N F_{\al}^{(i)}$ are operators on the tensor product module, 
defined by (\ref{X}).
The generalized Gaudin Hamiltonians commute between themselves, but only $N-1$ of them 
are independent.
Due to the property (\ref{P1}) we can complete this system by other 
$r$ operators,
the generators $\HH_{\al}$ of the Cartan sub-algebra, which commute with all ${\cal H}_i$.
For this family of independent and commuting Hamiltonians there is a
complete system of common eigenvectors in $\Omega$.

%------------------------------------------------------------
\subsection{The structure of $\Omega$}
%------------------------------------------------------------

Since the space $\Omega$ is a tensor product of $N$ highest weight 
representations $V_{\la}$ of $G$, it is useful to recall 
some basic results concerning the theory of highest weight representations of Lie algebras. 
 
Such a representation is completely determined by a highest weight vector 
$v_{\la}$, on which the action of the algebra is given by
\[
H_{\al} v_{\la}=<\al ,\la> v_{\la},\qquad
E_{\al} v_{\la}=0.
\]
As in the $SL(2)$ case, 
the representation space $V_{\la}$ is linearly generated only by monomials in generators 
of negative roots:
\begin{equation}
\{v_{\al_1...\al_l}^{n_1...n_l}=F_{\al_1}^{n_1}...F_{\al_l}^{n_l}\ v_{\la}\}_
{\al_i \in\Delta_+,\ n_i\in {\mathbb{N}}},
\label{Fvgen}
\end{equation}
but in this case, the monomials are ordered with respect to the roots, as showed by the 
Poincar\'e-Birkhoff-Witt theorem.\cite{Wan}
These vectors are eigenvectors for the Cartan generators: 
\begin{equation}
H_{\al} v_{\al_1...\al_l}^{n_1...n_l}=\left(<\al,\la> -
\sum_{\beta\in\Delta_{+}} n_{\beta}<\al,\beta>\right) v_{\al_1...\al_l}^{n_1...n_l},
\label{Hvgen}
\end{equation}
and the action of a positive root generator on monomials $ v_{\al}^{n_\al}$ is
\begin{equation}
E_{\al} v_{\al}^{n_\al}=n_{\al}
\left( 2\frac{<\la,\al>}{<\al,\al>} - n_{\al}+1 \right)
v_{\al}^{n_{\al}-1}. 
\label{Evgen}
\end{equation}

Concerning the dimension of a highest weight representation of $G$, 
recall that a weight $\la$ is a dominant integral if
\[
r_{\al}=2\,\frac{<\la,\al>}{<\al,\al>}\in {\mathbb{N}},\qquad 
\forall\,\al\in\Delta_0.
\] 
If the highest weight $\la$ is not dominant integral, then the representation 
is infinite dimensional and irreducible.
If the highest weight $\la$ is dominant integral then there is an 
invariant subspace,\cite{Wan} generated by vectors 
$v_{\al_1...\al_l}^{n_1...n_l}$, with 
$n_{\al}\geq r_{\al}+1$.
The quotient representation, denoted  $V_{\la}$ is an irreducible, finite-dimensional 
representation of $G$, generated by ordered monomials: 
\[
\{v_{\al_1...\al_l}^{n_1...n_l}=F_{\al_1}^{n_1}...F_{\al_l}^{n_l}\ v_{\la}\}_
{\al_j \in\Delta_+,\ n_j=0,...,r_{\al_j}}.
\]
As in the $SL(2)$ case, the space $\Omega$   
is a finite-dimensional tensor product module of $G$, 
completely determined by the vacuum vector $v_0=v_{\la_1}\otimes ...\otimes v_{\la_N}$  
and generated by ordered monomials:
\[
%v_{\al_1...\al_m}^{(k_1...k_m)}=
F_{\al_1}^{(k_1)} F_{\al_2}^{(k_2)}...F_{\al_m}^{(k_m)} v_0,
\]
with $1\leq k_i\leq N$. 
Note that for finite-dimensional representations, $m$ can vary between 0 and a maximal value
$m_{\mbox{\tiny\rm max}}=\sum_{i=1}^N\sum_{\al\in\Delta_+} r_{\al}^i$. 
Note also that the action of the Hamiltonians on the vacuum vector is
\begin{eqnarray}
\HH_\al v_0&=&\left(\sum^N_{k=1}<\la_k,\al>\right) v_0,\qquad\al\in \Delta_0,\nonumber\\
\ch_i v_0&=&\left(\sum_{j=1,j\not=i}^N\frac{<\la_i,\la_j>}{z_i-z_j}\right) v_0,\qquad i=1,...,N.\nonumber
\end{eqnarray}

As in the case of the Lie algebra  $SL(2)$, the Cartan operators $\HH_{\beta}$ 
have the particular role to give a gradation of the tensor
product module:
\[
\Omega =\bigoplus {\cal V}_\mu,
\]
on  weight subspaces ${\cal V}_{\mu}$, of weight 
$\displaystyle{\mu=\sum_{i=1}^N \la_i - \sum_{j=1}^m \gamma_j}$,  
generated by $\HH_{\beta}$ eigenvectors:
\[
\HH_{\beta}\,F_{\gamma_1}^{(k_1)}...F_{\gamma_m}^{(k_m)} v_0=
\left( \sum_{i=1}^N <\la_i,\beta> - \sum_{j=1}^m <\gamma_j,\beta>\right)
F_{\gamma_1}^{(k_1)}...F_{\gamma_m}^{(k_m)} v_0,
\]
with $\gamma_1,...,\gamma_m\in\Delta_+$ and $1\leq k_i\leq N$.
Since a positive root can be written as a sum of simple roots, with 
positive integer coefficients, in a unique way, the weight subspaces will be 
labeled by a family of simple roots: ${\cal V}_\mu={\cal V}_{\al_1...\al_s}$, 
with $\al_1...\al_s\in \Delta_0$, not necessarily distinct and satisfying 
$\sum_{j=1}^m \gamma_j=\sum_{k=1}^s \al_k$. Hence
\[
\Omega\ =\ \bigoplus {\cal V}_{\al_1 ...\al_s}.
\]

From the explicit form of the generalized Gaudin Hamiltonians 
(\ref{Higen}) and 
from the action (\ref{Fvgen})-(\ref{Evgen}) 
of the Lie algebra on each module $V_{\la_i}$, 
it follows that the weight subspaces ${\cal V}_{\al_1...\al_s}$ 
are invariant under the 
action of any ${\cal H}_i$. Therefore, in each subspace ${\cal V}_{\al_1...\al_s}$,
we can construct a basis of common eigenvectors of ${\cal H}_i$.

Unlike the case of the Lie algebra $SL(2)$, there is no result 
concerning the dimension of the invariant  subspaces ${\cal V}_{\al_1...\al_s}$
and neither for Sing ${\cal V}_{\al_1...\al_s}$. However,
using the properties (\ref{P2}) and (\ref{P3}), the recursive procedure 
presented in Sec. III can be generalized to construct in each subspace 
${\cal V}_{\al_1...\al_s}$ a family of common nonsingular eigenvectors, 
by applying operators $\F_{\al_i}$ on vectors of the sub-spaces 
${\cal V}_{\al_1...\al_{i-1}\al_{i+1}...\al_s}$. 
Concerning the subspace of singular vectors in  ${\cal V}_{\al_1...\al_s}$
we do not have a generalization of the result presented in Sec. III A, for the 
construction of a basis of singular vectors. The only result \cite{BabujianFlume} concerning this problem is a generalization of the Bethe Ansatz, which we discuss hereafter.

\subsection{Common eigenvectors in ${\cal V}_{\al}$}

Consider a simple root $\al$. 
The subspace  ${\cal V}_{\al}$ of weight $\sum_{i=1}^N \la_i - \al$ 
is generated by vectors 
$\left\{F_{\al}^{(k)}  v_0 \right\}_{k=1,...,N}$.
A Bethe vector in this subspace is defined as an
expansion on the basis, with rational coefficients 
depending on one complex parameter $w$:
\begin{eqnarray}
\psi_1(w,\alpha)v_0=\sum_{i=1}^{N}\frac{F_{\alpha}^{(i)}}{w-z_i}v_0={\cal F}(w,\alpha)v_0.
\label{psi1}
\end{eqnarray}
To give the action of a Hamiltonian ${\cal H}_i$ on this vector it is 
useful to calculate the commutator:
\begin{eqnarray}
\left[{\cal H}_i, {\cal F}(w,\al)\right]&=&
{\cal F}(w,\al)\frac{H_{\al}^{(i)}}{w-z_i}-\frac{F_{\al}^{(i)}}{w-z_i}
\left(\sum_{j=1}^N \frac{H_{\al}^{(j)}}{w-z_j}\right)\nonumber\\
& &+\sum_{\beta\in\Delta_+,\beta > \al}\frac{<\beta,\beta>}{2}N_{\beta,-\al}
\left\{{\cal F}(w,\beta)\frac{E_{\beta -\al}^{(i)}}{w-z_i}-
\frac{F_{\beta}^{(i)}}{w-z_i}\sum_{j=1}^N \frac{E_{\beta -\al}^{(j)}}{w-z_j}
\right\}.
\label{Hpsial}
\end{eqnarray}
Applied on the vacuum vector all the terms of the last sum vanish and 
the action of a Gaudin Hamiltonian ${\cal H}_i$ on the vector (\ref{psi1}) is
\begin{eqnarray}
{\cal H}_i\psi_1(w,\al)v_0=\!\left(\sum_{j=1;j\not=i}^N\!\!\frac{<\la_i,\la_j>}{z_i -z_j}\!+
\!\frac{<\al ,\la_i>}{w -z_i}\right) \psi_1(w,\al)v_0\! -\!
\left( \sum_{k=1}^N\frac{<\al ,\la_k>}{w -z_k}\!\right)\!\frac{F_{\al}^{(i)}}{w-z_i}v_0.
\label{Hipsi1}
\end{eqnarray}
Hence we have the following.
\begin{lema}
Given $N$ complex numbers $\left\{z_i \right\}_{i=1,...,N}$ and 
dominant integral highest weights $\left\{\la_i \right\}_{i=1,...,N}$, 
the Bethe vector $\psi_1(w,\al)v_0$ is an eigenvector for 
all Gaudin Hamiltonians:
\[
{\cal H}_i\,\psi_1(w,\al)v_0=s_i^1(w,\al)\,\psi_1(w,\al)v_0,\qquad\forall\ i=1,...,N,
\]
if the complex parameter $w$ satisfies the Bethe equation:
\begin{equation}
\sum_{k=1}^N\frac{<\al ,\la_k>}{w -z_k}=0.
\label{betheal}
\end{equation}
The eigenvalue $s_i^1(w,\al)$ of the Hamiltonian ${\cal H}_i$ depends 
on the solution of this equation: 
\[
s_i^1(w,\al)=\sum_{j=1;j\not=i}^N\frac{<\la_i,\la_j>}{z_i -z_j}+
\frac{<\al ,\la_i>}{w -z_i},
\qquad i=1,...,N.
\]
\label{lemaa1}
\end{lema}

%-----------------------------

\subsection{Common eigenvectors in ${\cal V}_{\al_1,\al_2}$}

Consider two simple roots $\al_1$ and $\al_2$, which are not necessarily distinct. 
The subspace ${\cal V}_{{\al}_1,\al_2}$, of weight 
$\ \sum_{i=1}^N \la_i -\al_1-\al_2\ $ 
is generated by vectors 
$\left\{F_{\al_1}^{(k_1)} F_{\al_2}^{(k_2)} v_0 \right\}_{k_1, k_2=1,...,N;\ k_1\not=k_2}$,
with two generators of negative roots applied on two distinct components of $v_0$,
but also  by vectors 
$\left\{(F_{\al_1}F_{\al_2})^{(k)} v_0 \right\}_{k=1,...,N}$ and 
$\left\{(F_{\al_2}F_{\al_1})^{(k)} v_0 \right\}_{k=1,...,N}$ with two generators 
of negative roots applied on the same component of $v_0$. 
A Bethe vector in ${\cal V}_{\al_1,\al_2}$ is defined as 
an expansion on all these vectors, with some particular coefficients 
depending on two complex parameters $w_1, w_2$:
\begin{eqnarray}
\psi_2(w_1, \al_1;w_2, \al_2)v_0&=&
\sum_{k_1=1}^{N}\sum_{k_2\not=k_1}^{N}
\frac{F_{\al_1}^{(k_1)} F_{\al_2}^{(k_2)}}{(w_1-z_{k_1})(w_2-z_{k_2})} v_0\nonumber\\
&+&\frac{1}{w_1-w_2}\sum_{k=1}^{N}\frac{(F_{\al_1} F_{\al_2})^{(k)}}{w_2-z_{k}} v_0+
\frac{1}{w_2-w_1}\sum_{k=1}^{N}\frac{(F_{\al_2} F_{\al_1})^{(k)}}{w_1-z_{k}} v_0\nonumber.
\end{eqnarray}
As element of the representation space $V_{\la_k}$, one of the two last terms 
is not a good ordered 
monomial and we must write, for instance,  
$(F_{\al_2} F_{\al_1})^{(k)}=(F_{\al_1} F_{\al_2})^{(k)}+
\left[F_{\al_2}, F_{\al_1}\right]^{(k)}$.
Then this vector can be also written in the following 
form, using the operators ${\cal F}$:
\begin{equation}
\psi_2(w_1, \al_1;w_2, \al_2)v_0={\cal F}\left( w_1, \al_1\right) {\cal F}\left( w_2, \al_2\right) v_0+
\frac{1}{w_1-w_2}{\cal F}\left( w_1, \left[F_{\al_1},F_{\al_2}\right]\right) v_0.
\label{twopsi}
\end{equation}

In order to compute the action of a Hamiltonian ${\cal H}_i$ on the vector $\psi_2$,
we use the result (\ref{Hpsial}) and (\ref{Hipsi1}) to obtain
\begin{eqnarray}
{\cal H}_i\,\psi_2(w_1, \al_1;w_2, \al_2)v_0&=&
s_i^2(w_1, w_2)\ \psi_2(w_1, \al_1;w_2, \al_2)v_0 \nonumber\\ 
& &-f_1\left( \frac{F_{\al_1}^{(i)}}{w_1-z_i}{\cal F}(w_2,\al_2)+
\frac{1}{w_1-w_2}\frac{\left[F_{\al_1}^{(i)},F_{\al_2}^{(i)}\right]}{w_1-z_i}\right) 
v_0\nonumber\\
& &-f_2\left( {\cal F}(w_1,\al_1)\frac{F_{\al_2}^{(i)}}{w_2-z_i}+
\frac{1}{w_1-w_2}\frac{\left[F_{\al_1}^{(i)},F_{\al_2}^{(i)}\right]}{w_1-z_i}\right) v_0,\nonumber
\end{eqnarray}
where
\begin{eqnarray}
s_i^2&=&\sum_{j=1;j\not=i}^N\frac{<\la_i,\la_j>}{z_i -z_j}+
\frac{<\al_1, \la_i>}{w_1 -z_i} + \frac{<\al_2,\la_i>}{w_2 -z_i},\nonumber\\
f_1&=&\sum_{k=1}^N\frac{<\al_1, \la_k>}{w_1 -z_k}+
\frac{<\al_1,\al_2>}{w_2-w_1}\ \ \mbox{\rm and}\ \ 
f_2=\sum_{k=1}^N\frac{<\al_2, \la_k>}{w_2 -z_k}+
\frac{<\al_1,\al_2>}{w_1-w_2}.\nonumber
\end{eqnarray}
Note that the action of ${\cal H}_i$ on the second term of the vector (\ref{twopsi})
must be calculated  separately, because (\ref{Hipsi1}) does not hold for 
the positive root $\al_1+\al_2$ which is no more simple. 
It follows from these considerations the following lemma.
\begin{lema}
The Bethe vector $\psi_2(w_1, \al_1;w_2, \al_2)v_0$ 
is an eigenvector for all Gaudin Hamiltonians,
\[
{\cal H}_i\,\psi_2(w_1,w_2)v_0=s^2_i\,\psi_2(w_1,\al_1; w_2,\al_2)v_0,
\]
if the parameters $w_1, w_2$ satisfy the Bethe equations: 
\begin{equation}
f_1=\sum_{k=1}^N\frac{<\al_1, \la_k>}{w_1 -z_k}+\frac{<\al_1,\al_2>}{w_2-w_1}=0,
\label{betheeq2}\qquad
f_2=\sum_{k=1}^N\frac{<\al_2, \la_k>}{w_2 -z_k}+\frac{<\al_1,\al_2>}{w_1-w_2}=0.
\end{equation}
\end{lema}

%-----------------------------

\subsection{Common eigenvectors in ${\cal V}_{\al_1,...,\al_m}$}

For $m$ simple roots $\al_1,...,\al_m$, not necessarily distinct, 
the subspace ${\cal V}_{{\al}_1,...,\al_m}$ of weight 
$\ \sum_{i=1}^N \la_i -\sum_{k=1}^m \al_k\ $ 
is generated by ordered monomials 
$\ F_{\gamma_1}^{(k_1)}... F_{\gamma_n}^{(k_n)} v_0\ $ with  $k_1,..., k_n=1,...,N$ and 
$\gamma_1,..., \gamma_n$ positive roots with $\sum_{i=1}^n\gamma_i=\sum_{k=1}^m\al_k$. 
In this general case it is difficult to define an appropriate Bethe type vector, 
with rational coefficients on the basis of ${\cal V}_{\al_1,...,\al_m}$. 
In Ref.  \cite{BabujianFlume} a recursive procedure was proposed  to 
define such a vector:
\begin{eqnarray}
& &\psi_m(w_1,\al_1;...;w_m,\al_m)v_0=\psi_{m-1}(w_1,\al_1;...;w_{m-1},\al_{m-1})
{\cal F}\left( w_m,\al_m\right) v_0\nonumber\\
& &+\sum_{j=1}^{m-1}\frac{1}{w_j-w_m}
\psi_{m-1}\left( w_1,\al_1;...;w_j,\left[F_{\al_j},F_{\al_m}\right];...;w_{m-1},\al_{m-1}\right) v_0,
\label{eq24}
\end{eqnarray}
with $\psi_1(w,\al)={\cal F}\left( w,\al\right)$.  
However, to our knowledge there is no proof that the action of a Hamiltonian 
${\cal H}_i$ on the vector $\psi_m v_0$ is
\begin{eqnarray}
& &{\cal H}_i\,\psi_m(w_1,\al_1;...;w_m, \al_m)v_0=
s_i^m\ \psi_m(w_1,\al_1;...;w_m,\al_m)v_0 -\sum_{k=1}^m f_k^m  \bar{\psi}_m^k 
%\frac{F_{\al_k}^{(i)}\psi_{m-1}(w_1,\al_1;...;w_{k-1},\al_{k-1};w_{k+1},\
%al_{k+1};...w_{m},\al_{m})}{w_k-z_i} 
v_0,
\label{eq25}
\end{eqnarray}
with  $\bar{\psi}_m^k$ some vectors in ${\cal V}_{{\al}_1,...,\al_m}$ and 
\begin{eqnarray}
s_i^m&=&\sum_{j=1;j\not=i}^N\frac{<\la_i,\la_j>}{z_i -z_j}+
\sum_{k=1}^{m}\frac{<\al_k, \la_i>}{w_k -z_i},\nonumber\\
f_k^m&=&\sum_{j=1}^N\frac{<\al_k, \la_j>}{w_k -z_j}+
\sum_{l=1,l\not=k}^{m}\frac{<\al_k,\al_l>}{w_l-w_k}.\nonumber
\end{eqnarray}

Such a recursive construction of generalized Bethe vectors seems to be appropriate 
for inductive calculations, but  
for the action (\ref{eq25}) of Hamiltonians on 
these vectors such a calculation  raises some problems, as explained in Ref. \cite{Eu2}.
Similar problems occur if we intend to prove inductively that all Bethe eigenvectors 
are singular. For a small number of roots this can be done by direct calculation.

We call $v^s\in \Omega$ a singular vector of $\Omega$ if all the generators 
of simple roots $\E_{\al}$ act trivially on $v^s$:
\[
\E_{\al}\, v^s\ =\ 0\,.
\]

Note first that $\left[E_{\beta},F_{\al}\right]=
\delta_{\al\beta}\frac{2}{<\al ,\al>} H_{\al}$ for any simple roots $\al$ and $\beta$.

Consider now the weight subspace ${\cal V}_{\al}$. 
If $\beta$ is a simple root we have
\begin{equation}
\left[\E_{\beta},\psi_1(w,\al)\right]=\delta_{\al\beta}\,\frac{2}{<\al,\al>}\,
\sum_{i=1}^{N}\frac{H_{\alpha}^{(i)}}{w-z_i}.
\label{Epsi1wa}
\end{equation}
Hence,
\[
\E_{\beta}\psi_1(w,\al)v_0=\delta_{\al\beta}\,\frac{2}{<\al,\al>}\,
\left( \sum_{i=1}^{N}\frac{<\alpha,\la_i>}{w-z_i}\right) v_0,
\qquad\forall\,\beta\in\Delta_0.
\]
Therefore, if $w$ is a solution of the Bethe equation (\ref{betheal}), then 
$\psi_1(w,\al)\,v_0$ is a singular vector of ${\cal V}_{\al}$.

In order to calculate the action of a generator $E_{\beta}$ of simple root
on a generalized Bethe vector (\ref{twopsi}) of ${\cal V}_{\al_1,\al_2}$, 
we use (\ref{Epsi1wa}) but also
\[
\left[\E_{\beta},\psi_1(w,\left[F_{\al_1},F_{\al_2}\right])\right]=
2\,\frac{<\al_1,\al_2>}{<\beta,\beta>}\,
\left\{\delta_{\beta\al_2}\,{\cal F}(w,\al_1)- 
\delta_{\beta\al_1}\,{\cal F}(w,\al_2)\right\},
\]
to obtain
\begin{eqnarray}
\E_{\beta}\psi_2(w_1,\al_1;w_2,\al_2)v_0&=&\delta_{\beta\al_1}
\frac{2}{<\al_1,\al_1>}\left\{\sum_{i=1}^N\frac{<\la_i,\al_1>}{w_1-z_i}+
\frac{<\al_1,\al_2>}{w_2-w_1}\right\}{\cal F}(w_2,\al_2)v_0\nonumber\\[0.5cm]
& &+\delta_{\beta\al_2}\frac{2}{<\al_2,\al_2>}\left\{\sum_{i=1}^N\frac{<\la_i,\al_2>}{w_2-z_i}+
\frac{<\al_1,\al_2>}{w_1-w_2}\right\}{\cal F}(w_1,\al_1)v_0.\nonumber
\end{eqnarray}
Therefore, if $(w_1,w_2)$ is a solution of the Bethe equations (\ref{betheeq2}), then 
$\psi_2(w_1,\al_1;w_2,\al_2)\,v_0$ is a singular vector of ${\cal V}_{\al_1,\al_2}$.

For Bethe vectors (\ref{eq24}) depending on more than 3 simple roots we claim that the 
commutator of $\E_{\beta}$ with $\psi_m$ is given by
\begin{eqnarray}
\left[\E_{\beta},\psi_m(w_1,\al_1;...;w_m,\al_m)\right]&=&
\sum_{i=1}^m \delta_{\beta\al_i}\frac{2}{<\al_i,\al_i>}\left\{
\psi_{m-1}(...;\widehat{w_i,\al_i};...)
\sum_{j=1}^{N}\frac{H_{\alpha_i}^{(j)}}{w_i-z_j}
\phantom{\sum_{j=1,j\not=i}^m}\right.\nonumber\\
& &-\left.\left(\sum_{j=1,j\not=i}^m\frac{<\al_i,\al_j>}{w_i-w_j}\right)
\psi_{m-1}(...\widehat{w_i,\al_i}...)\right\},\nonumber
\end{eqnarray}
where $\psi_{m-1}(...\widehat{w_i,\al_i}...)$ denotes 
$\psi_{m-1}(...w_{i-1},\al_{i-1};w_{i+1},\al_{i+1}...)$.
As for the action (\ref{eq25}) of the Hamiltonians, there is no inductive proof of this 
relation. 
The action of $\E_{\beta}$ on the Bethe vector 
is then
\[
\E_{\beta}\psi_m(w_1,\al_1;...;w_m,\al_m)v_0=
\frac{2\delta_{\beta\al_i}}{<\!\beta,\beta\!>}\left\{\!
\sum_{j=1}^N\frac{<\!\al_i,\la_j\!>}{w_i-z_j}-\!\!\!\!
\sum_{\mbox{\scriptsize{$\begin{array}{l}
k=1\\k\not=i\end{array}$}}}^m
\!\!\!\frac{<\!\al_i,\al_k\!>}{w_i-w_k}\!\right\}
\psi_{m-1}(...\widehat{w_i,\al_i}...)v_0.
\] 
Hence, the Bethe vector is singular 
if the parameters $w$ satisfy the Bethe equations $f_k^m=0$, $\forall\ k=1,...,m$.
%\newpage

\section{Conclusions}

In this article we recall the Gaudin model associated to the Lie algebra $SL(2)$ 
and its partial diagonalization by the Bethe Ansatz.

We give a general recursive method to construct
a  basis of common eigenvectors in each invariant subspace ${\cal V}_m$.
Knowing a basis of ${\cal V}_{m-1}$, we construct a family of nonsingular
independent common eigenvectors in ${\cal V}_m$.
This family is completed to a basis of ${\cal V}_m$, by a basis of the 
subspace of singular vectors of ${\cal V}_m$.

In order to describe the subspace Sing ${\cal V}_m$, we establish a relation
between singular vectors and covariant differential operators. This allows us 
to construct a basis of Sing ${\cal V}_m$ using an analog of the Gordan operator.
On the other hand we show that the Bethe Ansatz gives a family of singular common eigenvectors.
If Bethe equations have a maximal number of distinct solutions, then 
Bethe eigenvectors form a basis of common eigenvectors in  Sing ${\cal V}_m$. 

We discuss also the generalization of this method to the case of an arbitrary Lie algebra.
We recall the generalization of the Gaudin Hamiltonians  
and of the Bethe Ansatz. The generalized Bethe vectors are defined recursively, 
but this definition allows neither to calculate the action of the Hamiltonians,
nor to prove that Bethe eigenvectors are singular.
For a small number of simple roots we prove by direct calculation that the Bethe 
equations are conditions that Bethe type vectors be singular and common eigenvectors.
The recursive method to construct nonsingular common eigenvectors could also 
be generalized to the case of an arbitrary Lie algebra.

\section*{Acknowledgment}
We thank Prof. Richard Grimm for his comments on the manuscript and constant support.


\begin{thebibliography}{99}

\bibitem{Gaudinart}
\newblock {M. Gaudin},
\newblock {``Diagonalisation d'une classe d'hamiltoniens de spin''},
\newblock {\em Le Journal de Physique}, {\bf 37} (10), 1087--1098, 1976.

\bibitem{Gaudin}
{M. Gaudin},
\newblock {\em {La fonction d'onde de Bethe}},
\newblock S\'erie Scientifique, Masson, Paris, 1983.

\bibitem{FFR}
{B.L. Feigin, E. Frenkel and N. Reshetikhin},
\newblock {``Gaudin model, Bethe Ansatz and correlation functions at the critical
  level''}, 
\newblock {\em {Commun. Math. Phys.}} {\bf 166}, 27--62, 1994.

\bibitem{BabujianFlume}
{H.M. Babujian and R. Flume},
\newblock {``Off-shell Bethe Ansatz equation for Gaudin magnets and solutions of
  Knizhnik-Zamolodchikov equations''},
\newblock {\em Mod. Phys. Lett.} A {\bf 9}, 2029--2040, 1994.

\bibitem{Babujian}
{H.M. Babujian},
\newblock {``Off-shell Bethe Ansatz equation and N-point correlators in the SU(2)
  WZNW theory''},
\newblock {\em J. Phys.} A {\bf 26}, 6981--6990, 1993.

\bibitem{VarchenkoSchechtman}
{V.V. Schechtman and A.N. Varchenko},
\newblock {``Arrangements of hyperplanes and Lie algebra homology''},
\newblock {\em Invent. Math.} {\bf 106}, 139--194, 1991.

\bibitem{KZ}
{V.G. Knizhnik and A.B. Zamolodchikov},
\newblock {``Current algebra and Wess-Zumino model in two dimensions''}
\newblock {\em Nucl. Phys.} B {\bf 247}, 83--103, 1984.

\bibitem{RV}
{N. Reshetikhin and A. Varchenko},
\newblock {``Quasiclassical asymptotics of solutions to the KZ equations''},
\newblock {\em Preprint}, hep-th/9402126, 1994.

\bibitem{Go}
{P. Gordan}
\newblock {\em Invariantentheorie},
\newblock Chelsea Publishing Company, New York, 1987.

\bibitem{Eu2}
{D. Garajeu},
\newblock {``Generalizations of Gaudin Model and their relations with
  Knizhnik-Zamolodchikov equations''},
\newblock {\em Preprint CPT}, P.4006, 2000.

\bibitem{Etingof}
{P.I. Etingof, I.B. Frenkel and A.A. Kirillov},
\newblock {\em Lectures on representation theory and Knizhnik-Zamolodchikov
  equations}, volume~58 of {\em Mathematical Surveys and Monographs}.
\newblock A. M. S., 1998.

\bibitem{Eu1}
{D. Garajeu},
\newblock {``Conformally and projective covariant differential operators''},
\newblock {\em Lett. Math. Phys.} {\bf 47}, 293--306, 1999.

\bibitem{Wan}
{Z.-X. Wan},
\newblock {\em {Lie Algebras}}.
\newblock {International Series of Monographs in Pure and Applied Mathematics},
  Pergamon Press, New York, 1975.

\end{thebibliography}
\end{document}